\documentclass[journal]{IEEEtran}
\usepackage{cite}

\ifCLASSINFOpdf
  \usepackage[pdftex]{graphicx}
\else
\fi
\usepackage{amsmath}
\usepackage{xcolor}
\usepackage{multirow}
\usepackage{amssymb}
\usepackage{float}
\usepackage{subfig}

\usepackage{url}

\begin{document}
%
\title{A Novel RF-assisted-Strobe System for Unobtrusive Vibration Detection of Machine Parts}
%
%
%

\author{~Dibyendu~Roy,~Arijit~Sinharay,~\IEEEmembership{Senior~Member,~IEEE},~Brojeshwar~Bhowmick,~\IEEEmembership{Senior Member,~IEEE},~Raj~Rakshit,~Tapas~Chakravarty,~\IEEEmembership{Senior Member,~IEEE},~and~Arpan~Pal,~\IEEEmembership{Senior Member,~IEEE}
\thanks{All of the authors are with TCS Research and Innovation, Kolkata, India, e-mail: roy.dibyendu@tcs.com, arijit.sinharay@tcs.com, b.bhowmick@tcs.com, raj.rakshit@tcs.com, tapas.chakravarty@tcs.com and arpan.pal@tcs.com.}}

\maketitle

\begin{abstract}
In this paper, we propose a novel non-contact vibration measurement system that is competent in estimating linear and/or rotational motions of machine parts. The technique combines microwave radar, standard camera, and optical strobe to capture vibrational or rotational motions in a relatively fast and affordable manner when compared to the current technologies. In particular, the proposed technique is capable of not only measuring common vibrational parameters (e.g. frequency, motor rpm, etc.) but also provides spatial information of the vibrational sources so that the origin of each vibrational point can be identified accurately. Furthermore, it can also capture the wobbling motion of the rotating shafts. Thus, the proposed method can find immense applications in preventive maintenance across various industries where heavy machinery needs to be monitored unobtrusively or there is a requirement for non-contact multi-point vibration measurement for any machine inspection applications.
\end{abstract}


\IEEEpeerreviewmaketitle

\section{Introduction}
\IEEEPARstart{C}{ondition} \textcolor{black}{monitoring plays a crucial role in heavy industries for tracking the health and performance of machines over time \cite{r1}, \cite{r2}. The vibrational and rotational frequencies are considered as the initial indicator to determine the signs of degradation of any machine. Thus, routinely monitoring these features helps in preventive maintenance for such machineries.}

\textcolor{black}{In the current scenario, the accelerometer sensors are widely used to measure the vibrational or rotational frequencies \cite{r3}, \cite{r3.1}, \cite{r3.2} for machine inspection purposes. These sensors are required to touch with the machine's body and hence, can measure the vibrations from the contact point only. In spite of the high measurement accuracy of the accelerometer sensors, it has the following concerns: First and foremost, if multiple points of a large vibrating surface are required to be monitored simultaneously, then we need to mount a number of sensors in each of the inspection point, thus making the setup costly and cumbersome \cite{r4}, \cite{r4.2}. Secondly, in some scenarios, it is not always possible to install sensors without mechanically loading the system, for example, measuring rpm of a drilling head. Installing sensors directly on the head may cause unwanted loading which in turn, affect the alignment of the head, thus, the performance of the system deteriorates over time. Moreover, attaching external sensors on the machine body, results in violation of insurance policies for high-cost machineries. Thirdly, the contact-based measurement often requires machine downtime for replacing or servicing the attached sensors \cite{r4.3} that significantly affects the production. Finally, this approach generally requires wiring of the on-board sensors to a central data acquisition system, making the vicinity of the machine more hazardous. Although connecting sensors through a wireless link may resolve the issue, such a scheme would require frequent charging of the hosted sensors that may cause inconvenience in reality. Thus, an unobtrusive measurement approach is preferred over the contact-based schemes to avoid the above mentioned drawbacks. Two of such existing approaches are 
Laser-Doppler-Vibrometer (LDV) \cite{r5} and video camera-based vibration measurement \cite{r6}. The LVD looks for a Doppler shift \cite{r5} in the optical frequency range to measure the vibrations of the target.  However, the principle behind this technique is constrained to a point measurement process. Hence, to inspect a large vibrational-surface, the laser requires performing a raster scan which generally takes huge time to complete, thus, may not be suitable in condition monitoring of machine in real-time \cite{r5}.
On the other hand, video camera-based vibration detection \cite{r6} requires the use of very high frame rate cameras to measure high-frequency vibrations/rotations. For example, to detect $60000$ rpm of a motor, it would be required to use at least $2000$ fps camera to satisfy the Nyquist criterion \cite{r6}. Such high fps cameras are very expensive \cite{r23} and require a considerable amount of computational power (to process such high frame rates) \cite{r24}, \cite{r25}. In addition, these cameras are susceptible to noise and have lower sensitivity, which may pose problems in low light or varying light scenarios \cite{r26}, \cite{r27}. Hence, this modality is generally not used in machine inspection use-cases.}

\textcolor{black}{To address the problems of non-contact vibration detection in an affordable and scalable manner, in our earlier works \cite{r7}, \cite{r7.1}, \cite{r7.2}, \cite{r7.3}, \cite{r7.4}, we have developed a software-controlled optical-stroboscope where high-frequency vibrations co-located at different spatial points can be detected by an ordinary video camera (frame rate of $30$ fps). In addition, we have separately investigated the possibility of the non-contact vibration detection based on the principle of radio-frequency (RF) microwave radar \cite{r8} because of its immense applications in detecting physiological signals (heart/ breathing rate) \cite{r9.2} as well as monitoring structural vibrations of large objects (such as building, bridge etc.) \cite{r9}, \cite{r9.1}. We also did study on 3D esitmation using camera for localising the desired point in 3D \cite{Broj2} \cite{Broj4} but that required motion of camera, and hence RF is practical.}

\textcolor{black}{In this work, we propose a combined \textit{RF-strobe} based sensing mechanism to estimate the vibrational frequencies from the reflected-RF Doppler spectrum and localize the corresponding sources with the stroboscope and camera. This strategy provides quick estimation of spatially-distributed frequencies in a multi-source scenario and can capture various important measurements related to the machine inspection such as vibrational frequencies, rotational speed, spindle wobbling, etc. Therefore, the main contributions of this paper are:}
\begin{enumerate}
\item \textcolor{black}{To sense vibration and rotational motion in a time-efficient manner, we have fused the traditional continuous-wave (CW) RF-radar with a camera associated stroboscopic system \cite{r7.1}. Based on the sensing information of the radar unit, the camera associated stroboscopic system can estimate and localize several vibrational/rotational sources autonomously with a measurement resolution of $\pm 1$ Hz.}

\item \textcolor{black}{To localize each vibrational source precisely and autonomously of a large vibrating surface, we have proposed a marker-less image processing algorithm that employs the traditional optical-flow technique and Principal Component Analysis (PCA) based motion estimation strategy. This criterion makes the system more practical to implement in real scenarios.}

\item \textcolor{black}{To sense and localize the source that exhibits wobbling motion because of faulty or miss-aligned rotors, we have utilized a pair of proposed sensing modules. Fusing the information from both units, we have successfully tracked the trajectory of the wobble motion with a mean error rate of $1.56\%$.}

\textcolor{black}{Finally, to validate the efficacies of the proposed technique and to justify our claims, extensive real-time experimentations have been performed and presented in this paper. We emphasize the fact that such an integrated study on the unobtrusive vibration sensing has not been reported till date.}
\end{enumerate}

The remainder of the paper is organized as follows: In Sec. II, we have briefly described our previous works on unobtrusive vibration detection and its limitations. The methodologies of the proposed system along with the marker-less image processing solution have been presented in Sec. III. In Sec. IV, we have introduced the basic details behind the wobbling motion generation in faulty rotors. The various experimental setups, to validate the sensing capabilities of the proposed system, have been addressed in Sec. V, followed by the results and discussions in Sec. VI. Finally, Sec. VII concludes our present work.

\section{Background}
In this section, we have briefly outlined our previous works on unobtrusive vibration detection through the camera-based optical stroboscope \cite{r7} and the Microwave RF-radar \cite{r8}.

\subsection{Camera-based Vibration Detection using Optical Strobing}
In camera-based vibration detection \cite{r6} technique, a series of image frames are required to be processed sequentially. However, measuring high-frequency vibrations accurately involves the use of specialized, expensive, very high frame rate (e.g., $1000$ fps) cameras \cite{r23}, to avoid the conventional 'aliasing' problem. 
In our previous work \cite{r7}, \cite{r7.1}, \cite{r7.2}, \cite{r7.3}, \cite{r7.4}, we have proposed an approach, for measuring such high-frequency vibrations using a commodity low-speed ($30$ fps) camera. This method utilizes an optical stroboscope that effectively modulates the vibration signal, shifting the frequency components within the Nyquist frequency \cite{r6} range of the camera. Specifically, due to the principle of optical strobing \cite{ r7.3}, when the strobing frequency is very close to the true frequency of vibration or sub-multiple of it, the vibrating surface appears to be stationary (or moving very slowly). Detecting such motionless or low-frequency signals in the captured video frames provides the clue for computing the actual vibrational frequency of the object. In this aspect, we have proposed two strategies \cite{r7}, \cite{r7.3} to determine the actual frequency of vibration.

In our initial work \cite{r7}, the strobing frequency has been increased iteratively until the unknown frequency of vibration is detected. This technique may not be practical as it requires a large number of iterations to converge into a solution.  The above work has been extended in \cite{r7.3} where pairwise relatively prime positive integer strobing frequencies are generated iteratively. Thus, for each strobing frequency, a simultaneous linear congruence has been obtained. After that, Chinese-Remainder-Theorem (CRT) \cite{r9.3} has been applied to estimate a unique solution of the unknown vibrational frequency utilizing the set of linear congruencies and the corresponding strobing frequencies.
This strategy significantly reduces the detection time over the linearly increasing strobing technique. However, it is still incompetent to detect multiple vibrations (generated from several sources) within a specific time limit.

In all the earlier mentioned cases, the duty cycle of the strobing pulse plays a vital role in obtaining accurate information about the unknown vibration. \textcolor{black}{To get the sub-multiple effect of the vibrating object, the duty cycle of the optical pulse should be as low as possible} \cite{r7.4}. Reducing the duty cycle creates low illumination, hence to enhance the illumination, the number of LEDs in the panel needs to be increased, which significantly expands the size of the LED panel, subsequently making the deployment challenging in practical scenarios.

To alleviate the above-mentioned challenge, in this work, we have introduced CW radar for measuring the vibrational frequencies that in turn guides the strobe unit to readily lock into the correct strobing frequencies. This hypothesis eliminates the need for analyzing any sub-multiple effect; hence, the requirement for low duty cycle strobing pulse is no longer required. 

\subsection{Microwave Radar based Vibration Detection}
The main objective of this paper is to estimate the vibrational frequencies quickly and accurately, and also localize the corresponding sources generating those vibrations. As the stroboscopic system consumes large time to sense vibrations, in this work, we have combined the traditional CW radar into it. The main reasons for choosing CW radar are because of its simplicity in circuit design (not necessary to design stable and jitter-free transmitting source), affordability, and low computation power. 

The received signal from a moving target contains all the information about the motion of the object. When this received signal is mixed with the same transmitted signal \cite{r8}, the return Doppler signal can easily be estimated, thus, from this signal, it is possible to infer the complex movements of the target. Unlike stroboscope, RF-based measurements are very fast and do not require any iterative approach to sense the vibration. Therefore, it offers an excellent opportunity to measure all the vibrations present in its Field-of-View (FOV) in a single measurement. Moreover, this technique can pick up every vibration present in its FOV (even for multiple vibrating objects) in a single shot. However, it cannot, in general, provide information about the locations of these vibrational sources. In other words, the Doppler spectrum doesn't provide any spatial information \cite{r9} (i.e. which frequency is coming from which part of the vibrating object in a multi-source scenario). So, fusing the above-mentioned modalities, namely \textit{RF-assisted-Strobe}, can deliver such spatial information as detailed in the next section.

\section{Methodology}
\begin{figure}[thbp]
\centering
\centerline{\includegraphics[height=5cm, width=6.5cm]{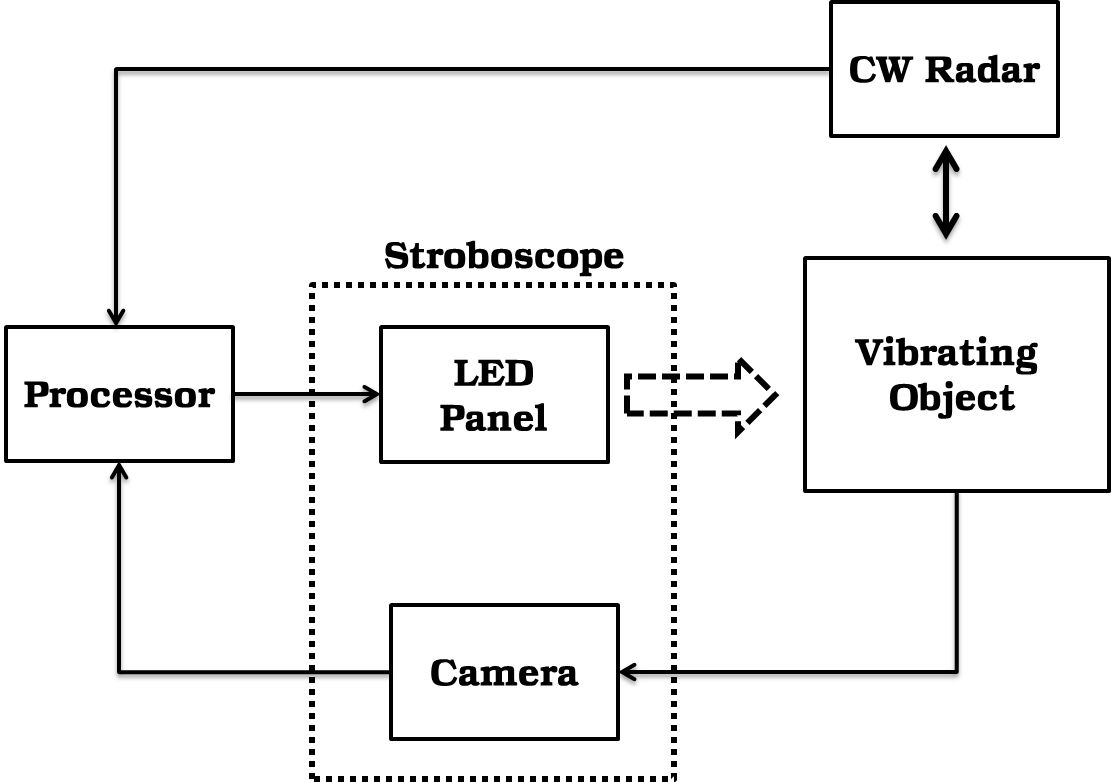}} 
\caption{\textcolor{black}{The hardware architecture of the RF-assisted-Strobe system.}}
\label{f2}
\end{figure}

In this section, we mainly describe the architecture of the RF-assisted-Strobe system. The overall sensing mechanism is represented in Fig. \ref{f2}.
The sensing process is divided into two parts: 1) RF-based sensing – to acquire the temporal frequency information, 2) Strobe-camera based sensing – to obtain the spatial information based on the temporal information of RF. Thus, the proposed sensing scheme is named as \textit{RF-assisted-Strobe}. 
RF can readily detect the vibrational frequencies in the Doppler spectrum that can serve as the triggering frequency for the strobe. Hence, the search space for the strobe is significantly reduced, thereby decreasing the iteration counts. Finally, the camera associated in the system will fine-tune the stroboscopic illumination to obtain spatial information of the vibrational frequency. The detailed description of the \textit{RF-assisted-Strobe} strategy is presented next.

\subsection{RF-assisted-Strobe Strategy}
This section provides the algorithmic steps of the proposed system. Before moving on to the detailed description, we have assumed that the object-of-interest (where multiple sources generating vibrations) lies within the FOV of the radar and camera. 
\begin{figure}[]
\centering
\includegraphics[width=8.5cm, height=7cm]{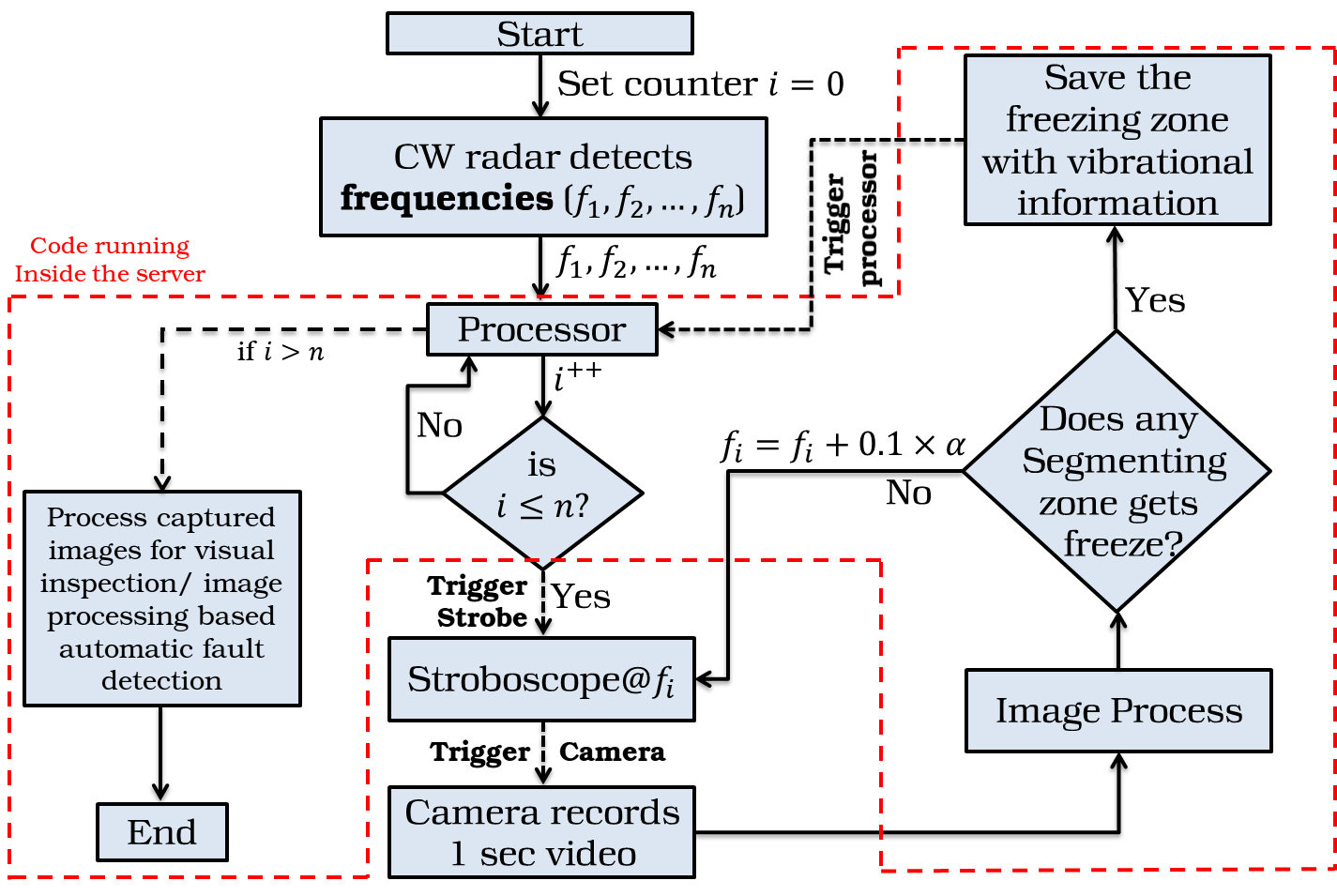}
\caption{\textcolor{black}{The algorithmic steps of the RF-assisted-Strobe sensing. \textit{Red}-dashed line indicates that this subsequent section is running inside the server.}}
\label{f9}
\end{figure}
At the beginning of the sensing scheme, the radar is triggered by the processor for some predefined duration. The return Doppler spectrum contains the gross-frequency values of $\begin{bmatrix}
f_1 & f_2 & \cdots & f_N
\end{bmatrix}$ in one shot, where $N$ represents the number of vibrating frequencies. This information is then passed on to the processor for triggering the stroboscope (i.e. the LED panel) to illuminate with the first frequency in the list, i.e. $f_1$. The source whose vibrating frequency nearest to $f_1$ would appear to be motionless or moving very slowly because of the strobing effect, thus, the resultant frames of that specific zone are within the Nyquist range of the camera. To determine the vibration of this section, the camera has been activated by the processor for some predefined duration. After capturing the images, the consecutive frames are analyzed to perceive the freezing effect of that particular region. While processing the consecutive frames, if the difference between the actual frequency (through the analysis of the subsequent frames) to that of the strobing frequency falls under $1$ Hz, then that corresponding zone appears to be motionless, subsequently, gets freeze. This condition serves as the stopping criterion to estimate the vibrational frequency of that unknown region. After detecting the vibration of that particular zone, the processor preserves this information and triggers the stroboscope to illuminate with a next frequency value of $f_2$. The entire procedure continues until it reaches $f_N$. In this aspect, we would like to mention that the gross frequency values as obtained from the radar can be fine-tuned using a parameter $\alpha$ to enhance the precision accuracy of the detected frequency. Lower values of $\alpha$ will increase the precision, convergence time and vice-versa. In our experiment, the typical range of $\alpha$ is $\{0\leq |\alpha| \leq 1\}$. 

In this work, our main objective is to estimate the frequency and rotation-per-minute (RPM) of rotational motions. To achieve this, we have applied our proposed system such that the advantages of each modality (i.e. RF, stroboscope, and camera) can be efficiently utilized. The detailed procedures comprising motion detection using radar, stroboscopic measurement and marker-less image processing techniques are explained in the subsequent sections.
\begin{figure}[thbp]
\centering
\centerline{\includegraphics[height = 4cm, width = 6cm]{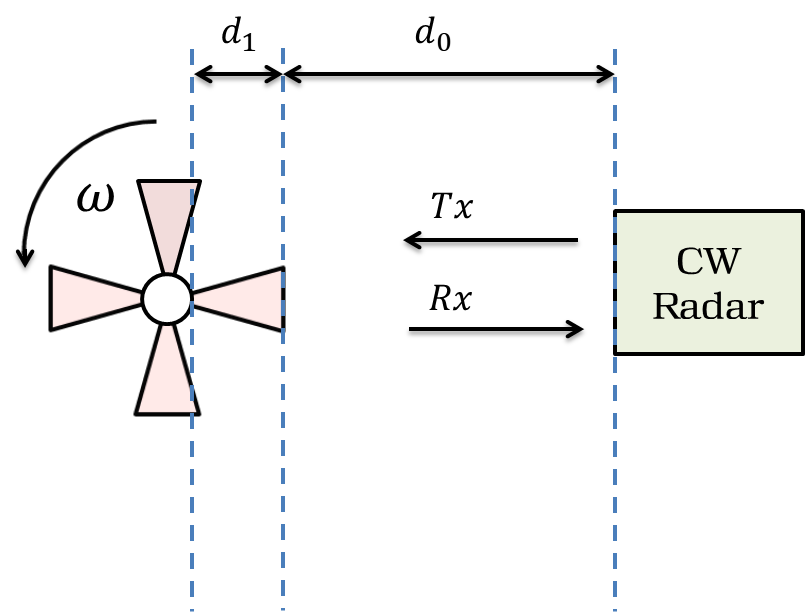}}
\caption{\textcolor{black}{Rotational motion detection using Microwave CW Radar.}}
\label{f3}
\end{figure}

\subsection{Motion Detection using Microwave CW Radar}
The CW-radars have traditionally been used to find the frequency \cite{r10} of one-dimensional periodic movement. However, in this work, a similar idea has been extrapolated to estimate the rotational motions of multiple rotating objects under the assumption that all the rotating objects are moving with different rotational speeds and they are operating within the FOV of the radar. For the CW-radar system, assuming $x(t)$ and $y(t)$ are the transmitted and received signals respectively. Then, the Doppler signal $b(t)$ can be represented as \cite{r10}, \cite{r11}
\begin{equation}
b(t)=\cos(\theta_o+\frac{4\pi w(t)}{\lambda}+\Delta\theta(t))
\label{e3}
\end{equation}
where $\lambda$ is the wavelength, $w(t)$ is the unknown periodic movement of the object, $\theta_o=\frac{4\pi d_0}{\lambda}$, $d_0$ is the distance between the radar and vibrating object. Eq. \ref{e3} exists if the motor blade hits the radial lines of RF excitation as depicted in Fig. \ref{f3}. The change in phase difference ($\Delta\theta(t)$) comes from the presence of blade vs. absence of the blade that provides the change in path length which in turn generates the Doppler signal. For example, if the RF wave radially hits the blades then the Tx to Rx path length becomes $2d_0$ whereas if it misses then the resultant path length becomes $2(d_1+d_0)$. This changing path length produces the Doppler signal which is modulated by the motor speed. Mathematically, the fundamental frequency of the Doppler signal is found to be at $(M\times \omega)$ rad/sec where $M$ is the total number of blades and $\omega$ is the rotational frequency (rad/s) of the motor \cite{r10}, \cite{r11}.

After computing the frequency spectrum of the Doppler signal, the processor will activate the stroboscope with a frequency value that has obtained from the Doppler spectrum.

\subsection{Motion Detection using Stroboscope}

\begin{figure}[!t]
\centering
\includegraphics[width=3in]{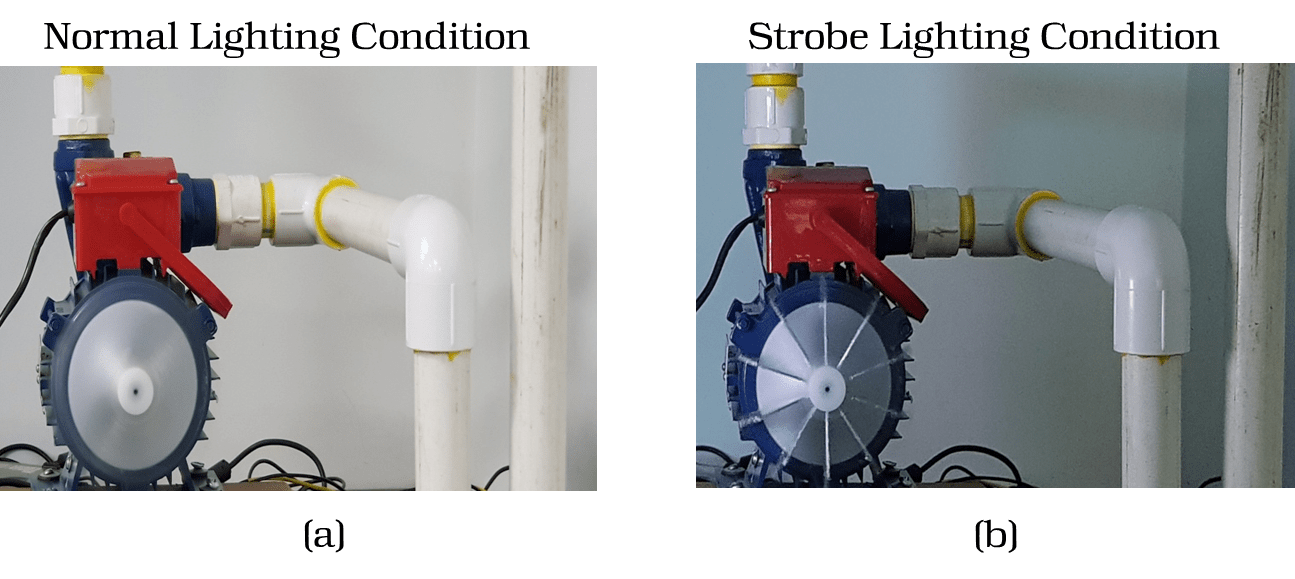}
\caption{Impact of Strobing on the Rotational Motion: (a) motion visualization in the ambient lighting condition, (b) motion visualization in the strobe light condition where the strobing frequency nearly matches with the motor RPM.}
\label{f4}
\end{figure}

Consider an object is rotating (or vibrating) with an unknown angular frequency (or vibrational frequency) of $\omega$ and the estimated frequency from radar is $\omega_s$. Therefore, as per the algorithm (vide in Sec. III-A), the stroboscope is triggered with a frequency of $\omega_s$. Thus, the rotational (or vibrational) motion of the object ($w(t)$) is optically sampled by an infinite pulse train (Dirac comb) $s(t)=\sum_{n=-\infty}^{\infty} \delta(t-nT_s)$ of known period $T_s=(2\pi)/\omega_s$. Then the resultant Fourier-spectrum of the sampled signal $W(\omega)$ will be
\begin{equation}
W(\omega)=\frac{1}{2\pi}\left[w(\omega)\circledast s(\omega_s) \right] = \frac{1}{T_{s}} \left[ W(\omega-\omega_{s})\right]
\label{e4}
\end{equation}
where $w(\omega)$ and $s(\omega_s)$ are the Fourier transform of the signal $w(t)$ and $s(t)$ respectively. The impact of optical illumination on the rotational motion (when applied on a domestic water-pump) is shown in Fig.~\ref{f4}. Fig.~\ref{f4}(a) represents a blurry vision of the rotating blades in the ambient lighting condition (i.e. without strobing). 
During strobing, when the strobing frequency ($\omega_{s}$) matches the true rotational frequency ($\omega$), the blades appear to be motionless (i.e. $||\omega_{s}-\omega||\leq 1$ Hz). One of such freezing effect is shown in Fig.~\ref{f4}(b).

In real scenarios, the strobing frequency ($\omega_s$) may not exactly match with $\omega$. So, to fine-tune $\omega_s$, image processing (on the captured video frames) could be further employed. Based on this consequence of the image processing step, $\omega_s$ is required to be adjusted for complete static frames so that $\omega_s\approx\omega$ \cite{r7.1}. 
A similar technique might also be very beneficial in a multi-source environment to sense each frequency component along with their corresponding source location. For example, whenever, a strobing frequency matches with any of the vibrational frequencies, that particular source section will seem to be static and easily distinguishable by the image processing step, thus providing the information about the source location.
To perform the image processing step, in our previous work \cite{r7.2}, a visible marker has been put on the vibrating surface so that the camera can track the motion to assess the static condition.
In this work, we have generalized the image processing by employing an optical-flow based methodology to eliminate the need for putting markers explicitly on the vibrating surface. The complete procedure is discussed in the next section.

\subsection{Marker-less Image processing}
\begin{figure}[!t]
\centering
\includegraphics[width=\columnwidth, height=3.2in]{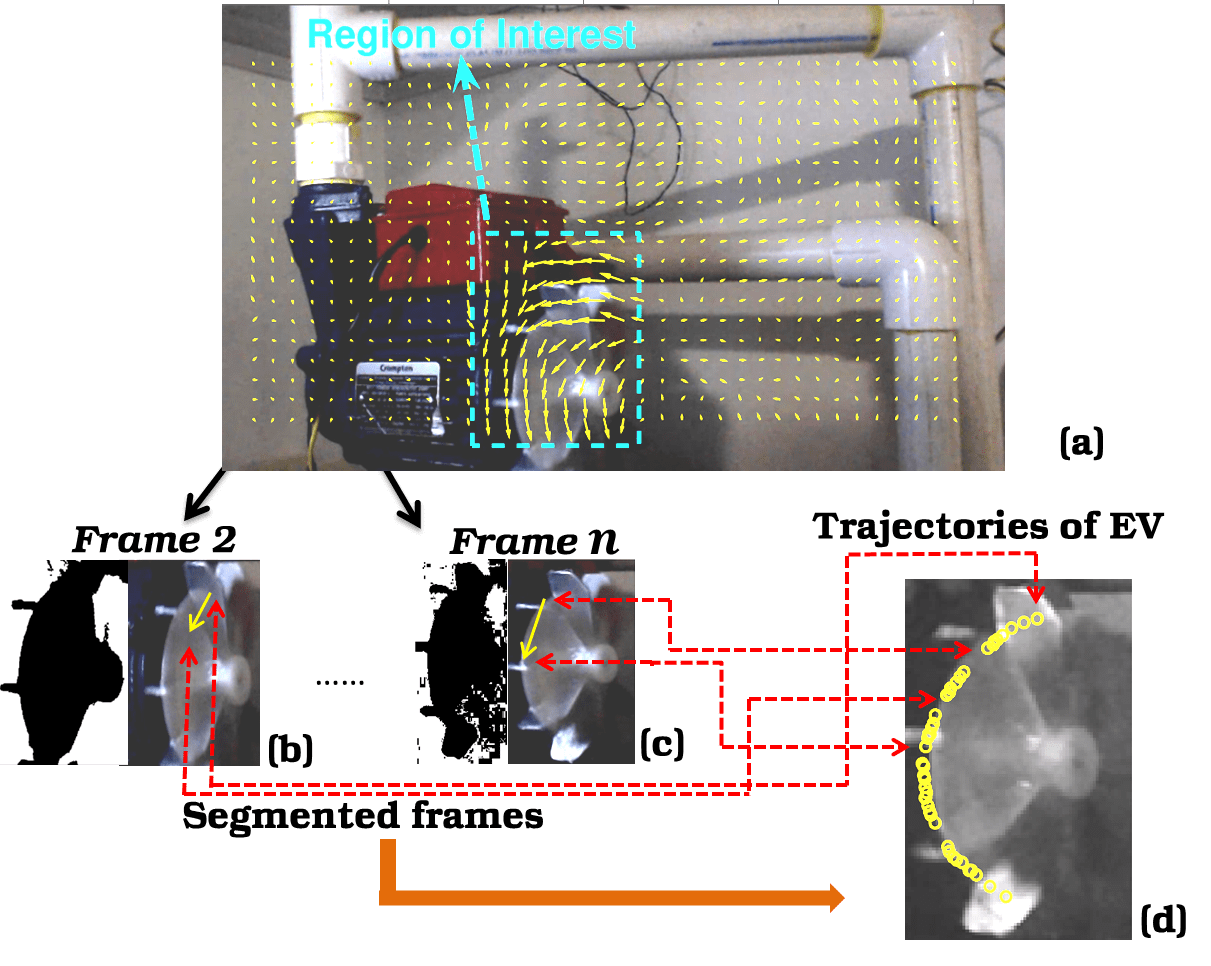}
\caption{The proposed marker-less image processing technique: (a) optical flow-based ROI segmentation, (b) $2^{nd}$ segmented frame with binary representation. The \textit{yellow}-arrow represents the eigenvector corresponding to the largest eigenvalue. (c) $n^{th}$ segmented frame with binary representation having the eigenvector corresponding to the largest eigenvalue. (d) Trajectories of the eigenvalues (EV) in each segmented frame. }
\label{f7}
\end{figure}

Our proposed marker-less image processing has two aspects: firstly, segmenting each region-of-interest (ROI) based on the motion estimation, and secondly, extracting vibrational information for each segmented zone. To achieve these objectives, we use the traditional optical flow technique \cite{r13} for ROI selection. The ROIs are segmented based on the motion information that has been obtained from the optical flow. One of such instance is shown in Fig. \ref{f7}(a). 
To obtain the vibrational or rotational information of each segmented region, we have analyzed the principal component (PCA) of the corresponding motion vector (as shown in Fig. \ref{f7}(d)).

The optical flow is the pattern of apparent motion of objects, surfaces, and edges in a visual scene caused by the relative motion between an observer and the scene \cite{r12}. Sequences of images allow the estimation of motion as either instantaneous image velocities or discrete image displacements. In this work, we have calculated the motion between two image frames which are taken at times $t$ and $t+\Delta t$ at every pixel position.
Assuming the $i^{th}$ pixel ($q_i$) at location $(x,y,t)\in \Re^{2+t}$ with intensity $I(x,y,t)$, is moved by $\Delta x$, $\Delta y$ and $\Delta t$ between two consecutive image frames. Then the brightness constancy constraint will be \cite{r12}
\begin{equation}
I(x,y,t)=I(x+\Delta x, y+\Delta y, t+\Delta t )
\label{e5}
\end{equation}
Then, the horizontal ($V_x$) and vertical ($V_y$) velocity components of the optical-flow can analytically be defined as \cite{r13}
\begin{equation}
\begin{split}
\begin{bmatrix}
V_x \\ V_y 
\end{bmatrix}
=
\begin{bmatrix}
\sum_i I_x (q_i)^2 & \sum_i I_x (q_i) I_y (q_i) \\
\sum_i I_y (q_i) I_x (q_i) & \sum_i I_x (q_i)^2 
\end{bmatrix}^{-1} \times \\
\begin{bmatrix}
\sum_i I_x (q_i) I_t (q_i) \\
\sum_i I_y (q_i) I_y (q_i) \\
\end{bmatrix}
\end{split}
\label{e10}
\end{equation}
where $I_x (q_i)$, $I_y (q_i)$ and $I_t (q_i)$ are the image intensity along $x$-axis, $y$-axis and $t^{th}$ time respectively for $q_i$ pixel. Therefore, equating Eq. \ref{e10}, a set of $(V_x,V_y)\forall[q_1,q_2,...,q_n]$ is obtained as shown in Fig. \ref{f7}(a).
After getting the velocity vectors for all the frames, we need to evaluate the features which will help to segment out the ROI. In this aspect, two basic image features are used: the image brightness ($L(x,y)$) and the Euclidean norm ($f(x,y)$) of the velocity vectors as in Eq. \ref{e11}
\begin{equation}
\begin{split}
L(x,y) &=  \begin{cases}
    1, & \text{if } I(x,y)\geq \lambda\\
    0,              & \text{otherwise}
\end{cases} \\
f(x,y) &= \sqrt{V_x^2 + V_y^2}
\end{split} \label{e11}
\end{equation}
where $\lambda$ is a threshold parameter for the image $I$. 
While $L(x,y)$ signifies the foreground points corresponding to brightness criteria, $f(x,y)$ denotes the points corresponding to the motion criteria. The points for which both the criteria meet is considered as the actual candidate points for the foreground regions. After that, the K-means clustering algorithm \cite{r15} is applied to the foreground motion features for accurate segmentation and localization. Figs. \ref{f7}(b) and \ref{f7}(c) represent the segmentation results of two different frames.

Now, in each segmented frame (considering the initial frame as reference), we have analyzed the principal-component \cite{r16} on the horizontal ($V_x$) and vertical ($V_y$) velocity vectors to obtain the orientation of the eigenvectors in \textit{2D-}space. The eigenvector corresponding to the largest eigenvalue is selected as the principal component because it indicates the direction of motion having maximum variance \cite{rLG}. In principle \cite{rLG2}, it provides the resultant direction of motion of the entire object by suppressing the contribution of the spurious motion, if any. The outcomes are shown in Fig. \ref{f7}(b) and \ref{f7}(c) respectively. Now, for each segmented frame, the eigenvectors indicate a trajectory of that corresponding part. If we accumulate all the eigenvectors for all the frames, we will get a holistic trajectory as shown in Fig. \ref{f7}(d).
Once, the trajectory has been obtained, we perform the Fourier analysis on the horizontal and vertical components of the resultant trajectory which will provide the strobing-modulated frequency content in the rotational motion. If this measured frequency meets the freezing criteria (as described in Sec. III-A), then the algorithm ends or performs a similar analysis for the next ROI. Otherwise, the additional frequency has been adjusted to the stroboscope by the processor to obtain the freezing effect.

\section{Wobbling motion Generation and Detection}
The method developed for detecting vibration or rotational motion can potentially be used to detect wobbling motion of faulty rotors. The wobbling provides an early signature for machine faults and thus may be a parameter of interest to be monitored \cite{r21}.
\begin{figure}[]
\centering
\includegraphics[width=3.5in, height=1.8in]{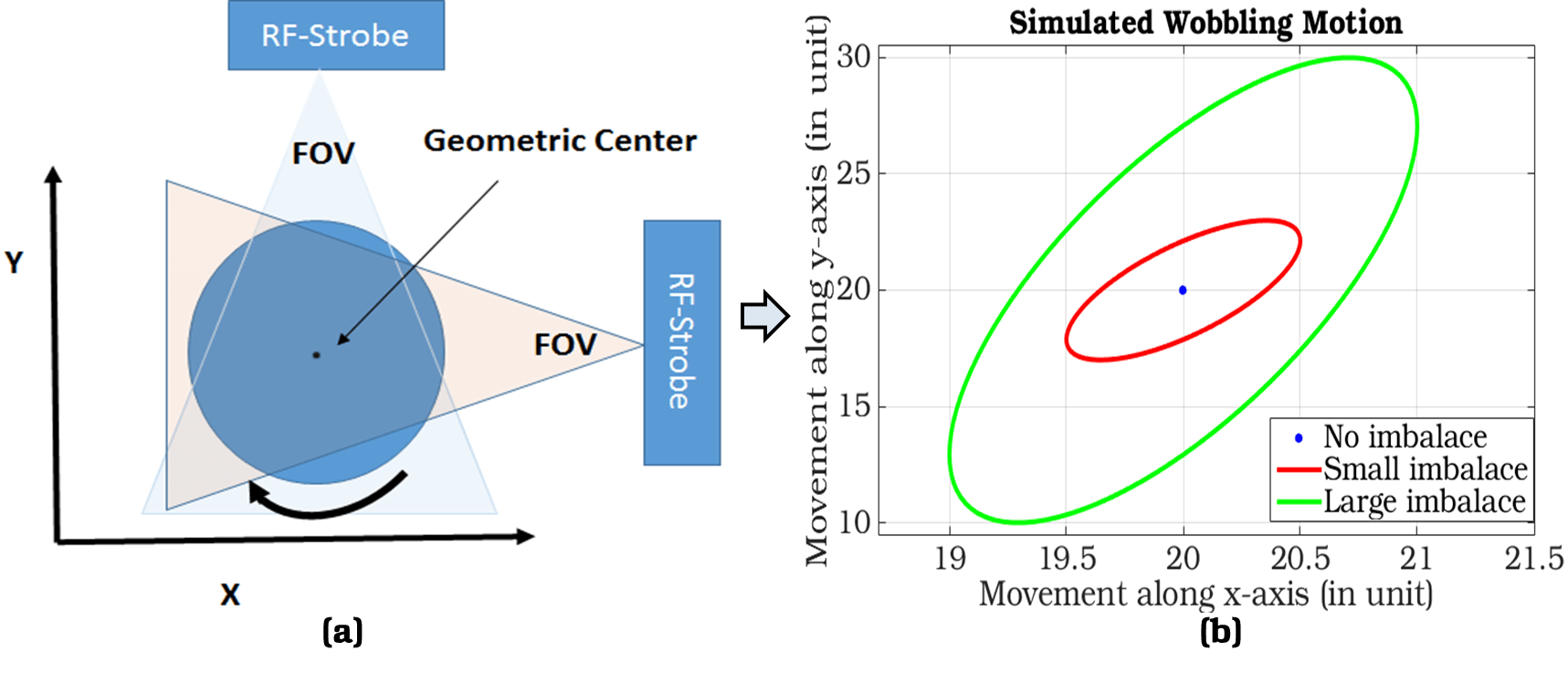}
\caption{\textcolor{black}{Wobbling motion detection,(a) Sensing principle, (b) Simulated trajectory}}
\label{f8}
\end{figure}
Wobbling happens when a rotating object suffers an imbalance \cite{r22}. For example, when a disc is balanced, it rotates around its geometric center (Fig \ref{f8}(a)). In this condition, no motion is seen along $X$ or $Y$ direction. During an imbalance situation, the center of mass shifts, and the geometric center revolves around the new center of mass. This aspect causes linear displacements in $X$ and $Y$ directions in a periodic fashion, thus, creating wobbling motion. Such kind of undesired motion can be sensed by the proposed system if the sensing is conducted from two orthogonal sides as depicted in Fig. \ref{f8}(a). The resultant trajectory in the $X-Y$ plane (after fusing $x$ and $y$ directional motion) may capture the nature of imbalance (Fig. \ref{f8}(b)). Analytically, such motion can be described as
\begin{equation}
\begin{split}
&x_1(t) = x_0 + A_x\cos(\omega_bt) \\
&y_1(t) = y_0 + A_y\cos(\omega_bt + \pi/2) = y_0 + A_y\sin(\omega_bt)
\label{12.1} 
\end{split}
\end{equation}
where $[x_1(t),y_1(t)]$ is the position of the center of mass of the system at $t^{th}$ time, $(x_0,y_0)$ is the initial location of the center of mass, $A_x$, $A_y$ are the wobble amplitude along $x$ and $y$ directions respectively, and $\omega_b$ is the wobble frequency (which is generally less than the motor rpm). In balanced condition, $A_x=A_y=0$. However, during an unbalanced situation, $A_x \neq A_y \geq 0 $, thus, $\omega_b>0$. Fig. \ref{f8}(b) shows the trajectories of a perfectly balanced rotating object (\textit{blue}-dot) and imbalanced rotating objects (\textit{green} and \textit{black} elliptical trajectories).

\section{Experimental Setup}

To validate the technique, several experiments have been performed with the industrial graded vibrating objects. \textcolor{black}{In this study, we have used the HB100 low-cost CW radar modules \cite{r8}, having an operating frequency of $10.5$ GHz (azimuth and elevation angles of $80^0$ and $40^0$ respectively). The detailed hardware architecture of this module can be found in \cite{r8}. It is approximately placed at $1$ meter distance from the object-of-interest, hence, the Field-of-view (FOV) is around $2.54\times 0.7$ $m^2$.}
For the stroboscopic hardware, a portable $20\times20$ white-light LEDs panel, controlled by an MSP432 \cite{r18} micro-controller board, is employed. In this aspect, we would like to mention that all the experiments have been conducted in a dark environment to eliminate the effect of the ambient light as the solution is very sensitive to natural light. The entire algorithm has been executed in an \textit{intel core i7}, $8$-GB processor with python and MATLAB 2015a software environments. Table \ref{table1} represents the parametric values used in this experiment. To reduce the sub-multiple effect of strobing, intentionally, we have increased the duty cycle of the strobing-pulse to $10\%$.
\begin{table}[thbp] 
\caption{Parametric values taken in the experiment } \vspace{-0.2cm}
\label{table1}
\centering
\begin{tabular}{|c|c|}
\hline
Parameter & Value \\
\hline
Duty Cycle & $10\%$ \\
\hline
\textcolor{black}{$\alpha$} & [\textcolor{black}{$-1$ to $1$}] \\
\hline
$\omega_c$ (camera frame rate) & $30$ fps \\
\hline
Radar "ON" time & $10$ sec. \\
\hline
Camera "ON" time & $1$ sec. \\
\hline
\end{tabular}
\end{table} 
\subsection{Exp 1: Multiple Linear Motion Detection}
\begin{figure}[]
\centering
\includegraphics[width=2.5in,height=2in]{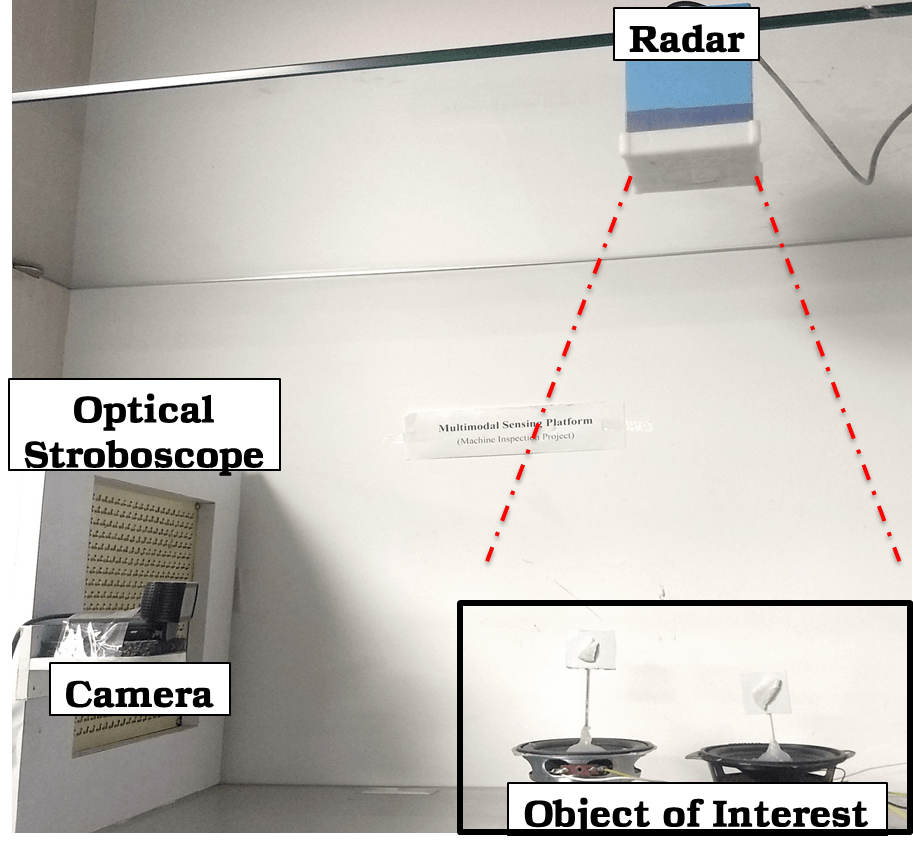}
\caption{\textcolor{black}{The RF-assisted-Strobe system for sensing multiple linear motion.}}
\label{f5}
\end{figure}

To detect linear vibration (i.e. vibration along a single-axis), we have used two traditional sound-speakers that are excited by a dual-channel signal-generator of known frequencies (as shown in Fig. \ref{f5}). The duration of "ON" time for the radar and camera is $10$ sec and $1$ sec respectively.  Moreover, to assess the freezing criteria from the captured frames, an irregular-shaped attachment has been connected to the diaphragm of each speaker. 

\begin{figure*}[t]
\centering
\subfloat[Singular Rotational Motion]{\includegraphics[width=2.2in,height=1.8in]{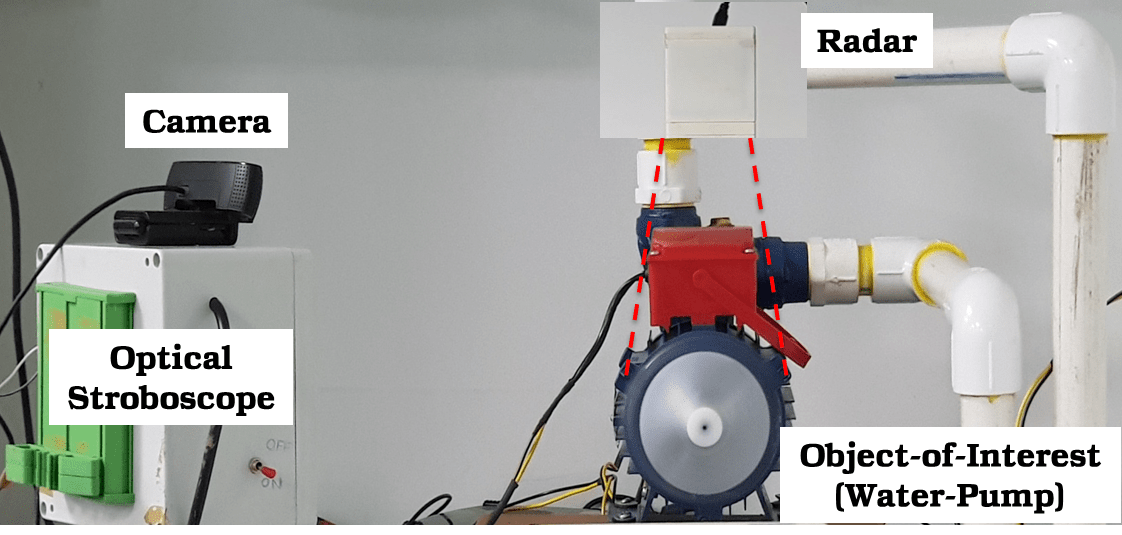}
\label{f10a}} 
\subfloat[Multiple Rotational Motion]{\includegraphics[width=2.3in,height=1.6in]{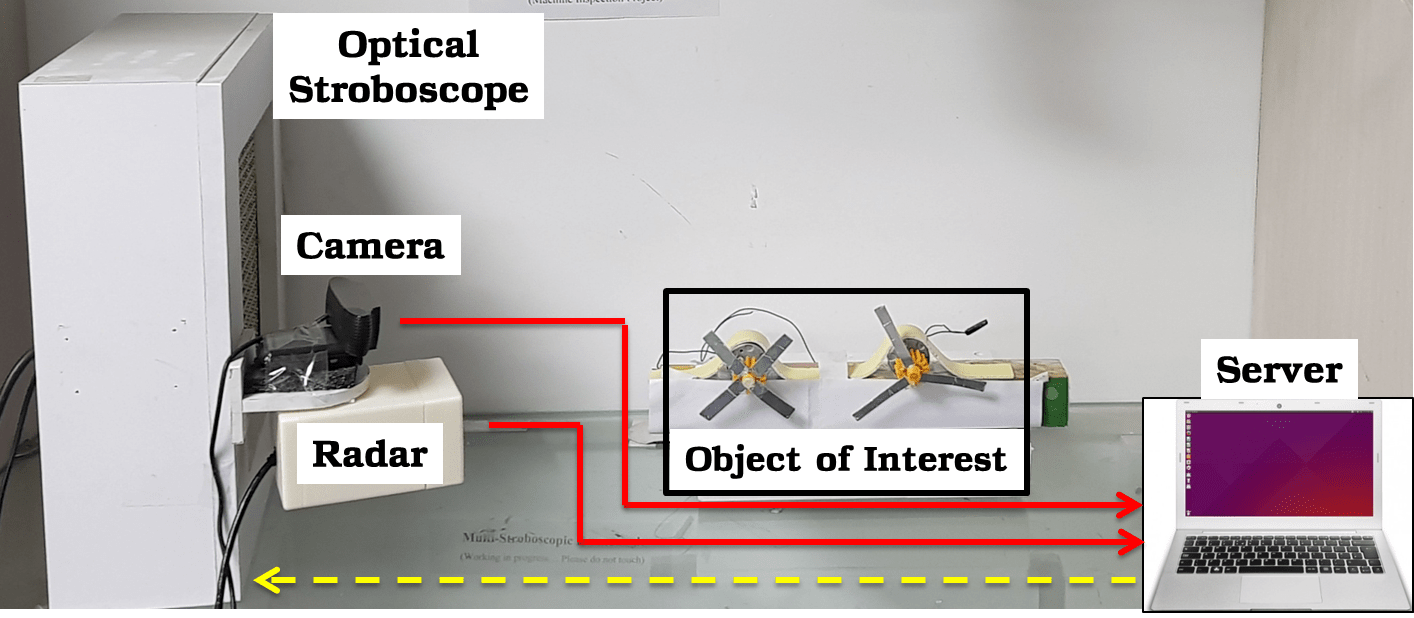}
\label{f10b}} 
\subfloat[Wobbling Motion]{\includegraphics[width=2.2in,height=1.8in]{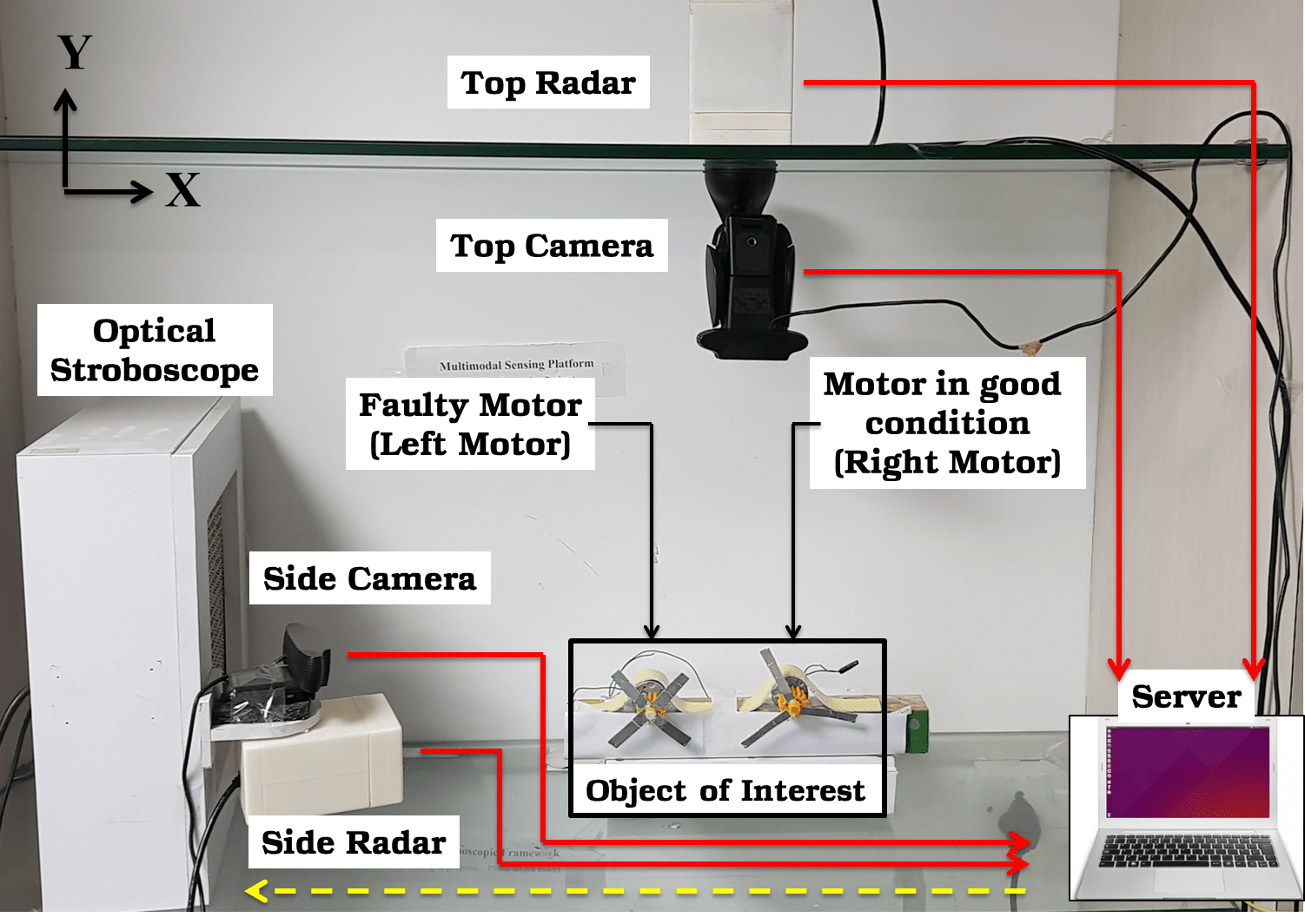}
\label{f10c}}
\caption{Experimental setup of the proposed system: (a) Rotational motion detection setup (b) Multiple rotational motion detection setup (c) Wobbling motion detection setup.}
\label{f10}
\end{figure*}
\subsection{Exp 2: Rotational Motion Detection}

To sense rotational motion, two different experiments are conducted: 1) single rotation motion sensing (shown in Fig. \ref{f10a}) with a domestic water-pump \cite{r20} (having an rpm of $2400$ with $10$ rotating blades), and 2) multiple rotational motion (shown in Fig. \ref{f10b}) sensing with a couple of motors (having a maximum rpm of $3600$ with $3$ and $4$ rotating blades respectively). To alter the rpm between these motors, a voltage-divider circuit has been added on the right motor (Fig. \ref{f10b}) so that it can revolve slower than the other. 
For both scenarios, the same sensing scheme (i.e. RF-assisted-strobe) has been applied. The radar is located at the vertically upward direction with-respect-to the rotational speed and the stroboscope associated with a camera is placed horizontally.

\subsection{Exp 3: Wobbling Motion Detection} 
For making the solution more versatile, we attempt to detect wobbling motion \cite{r19} of rotors so that it can be useful in real scenarios. To generate imbalance, fault has been injected in the left motor by untightening one screw (setup is shown in Fig. \ref{f10c}). As a result of this, the geometric center of the left motor starts rotating around the new center-of-mass (as described in Sec. IV). In this case, to sense the wobble motion, we have adopted a pair of camera and radar that are placed in the horizontal and vertical positions with-respect-to the rotational direction of the motors.

\section{Results and Discussions}
In this section, we have mainly discussed about the outcomes of the above-mentioned experimental setups.

\subsection{Results of Exp 1: Multiple Linear Motion Detection}
\begin{figure}[!t]
\centering
\includegraphics[width=\columnwidth,height=2.5in]{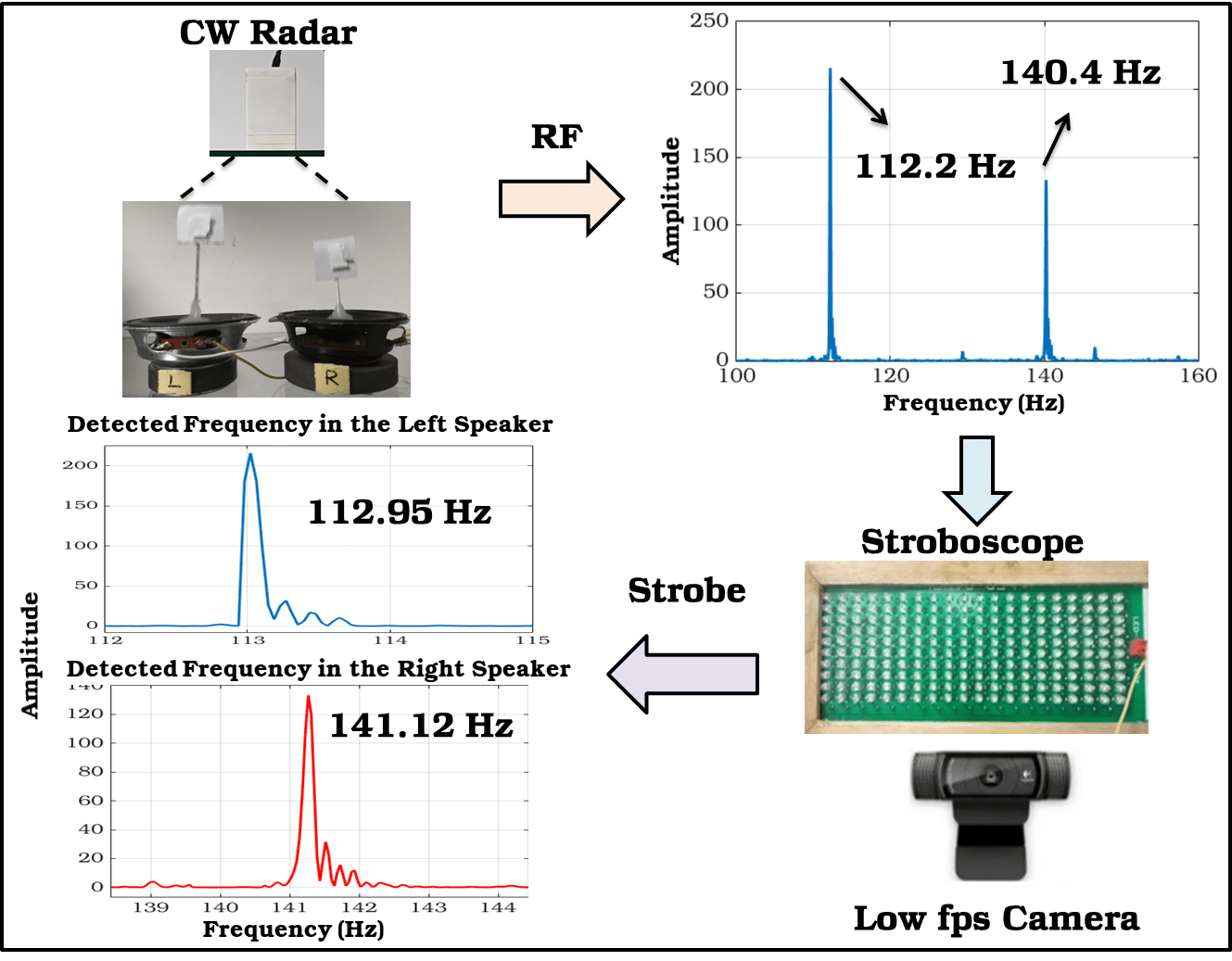}
\caption{Results of multiple linear motion setup with the proposed approach.}
\label{f11}
\end{figure}

In this experiment (as specified in Sec. V-A), the audio-speakers are excited with separate frequencies ($113$ Hz on the left-speaker and $141$ Hz on the right-speaker). The resultant Doppler-spectrum exhibits two frequency peaks at $112.2$ and $140.4$ Hz respectively as shown in Fig. \ref{f11}. Setting these frequency values in the stroboscope sequentially, the object-of-interest appears to vibrate slowly. The image processing is further applied to ensure the freezing criteria. 
While analyzing the captured frames (as discussed in Sec. III-D), $0.75$ and $0.72$ Hz peak frequencies are observed for the left and right speakers respectively. 
Adjusting these values into the stroboscope, the updated frequencies of vibration for the left and right speakers are obtained as $112.95$ and $141.12$ Hz respectively, subsequently generates a detection error of $0.04\%$ and $0.09\%$.
 \vspace{-0.6cm}
\subsection{Results of Exp 2: Single Rotational Motion Detection}
\begin{figure}[!t]
\centering
\includegraphics[width=\columnwidth,height=2.5in]{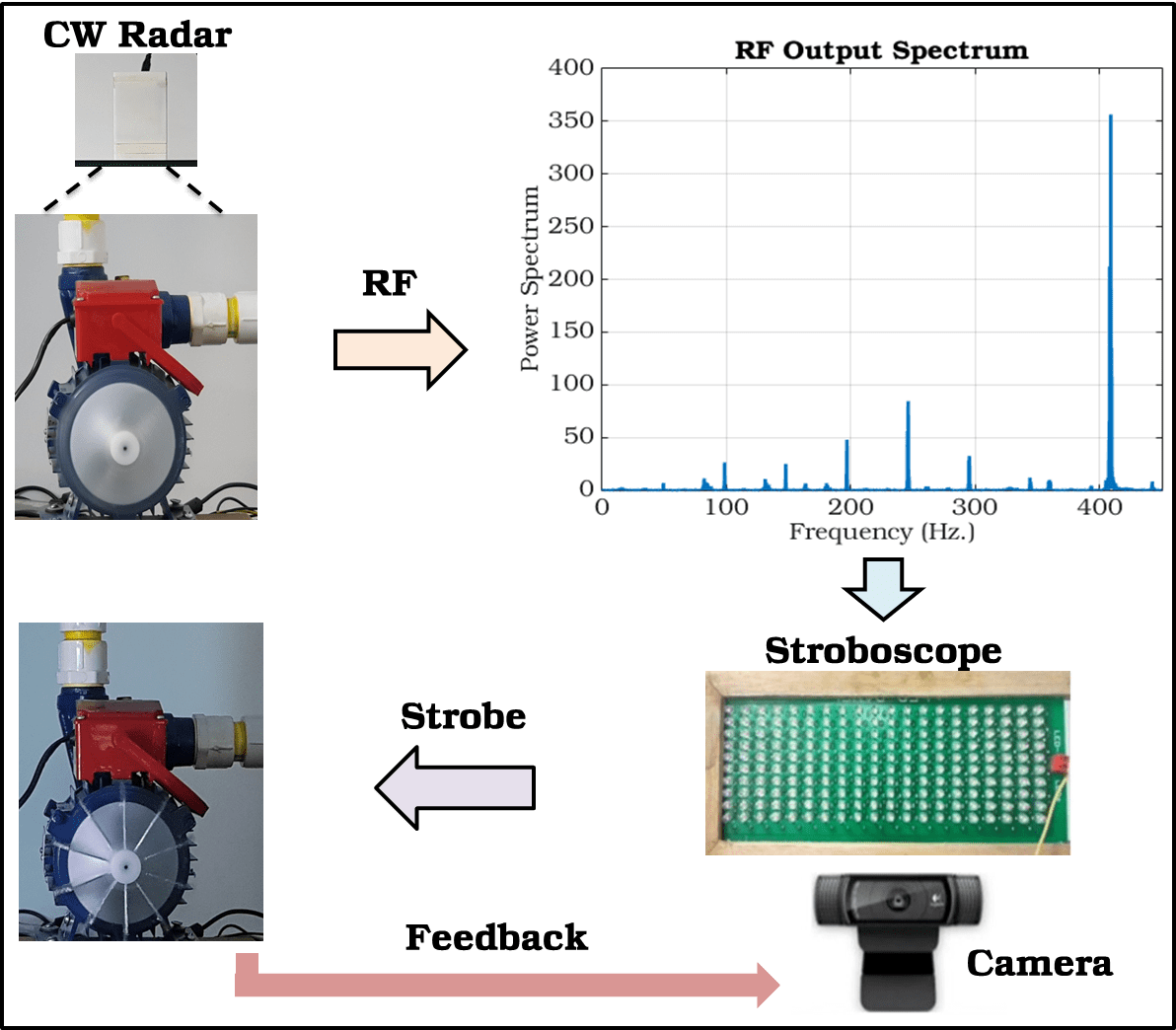}
\caption{Results of rpm measurement of an industrial-graded water-pump.}
\label{f12}
\end{figure}

In this study, the Doppler-spectrum of the radar shows a dominant peak frequency at $408.23$ Hz as shown in Fig. \ref{f12}. Hence the predicted fundamental frequency of rotation is approximately $40.82$ Hz (i.e. $408.23/10$; as the motor has $10$ blades). After fine-tuning the strobing frequency through the image processing steps, the final observed rotational frequency is $40.13$ Hz that corresponds to $2408$ rpm hence the detection error rate is $0.33\%$.

\begin{figure}[!t]
\centering
\includegraphics[width=\columnwidth,height=2.5in]{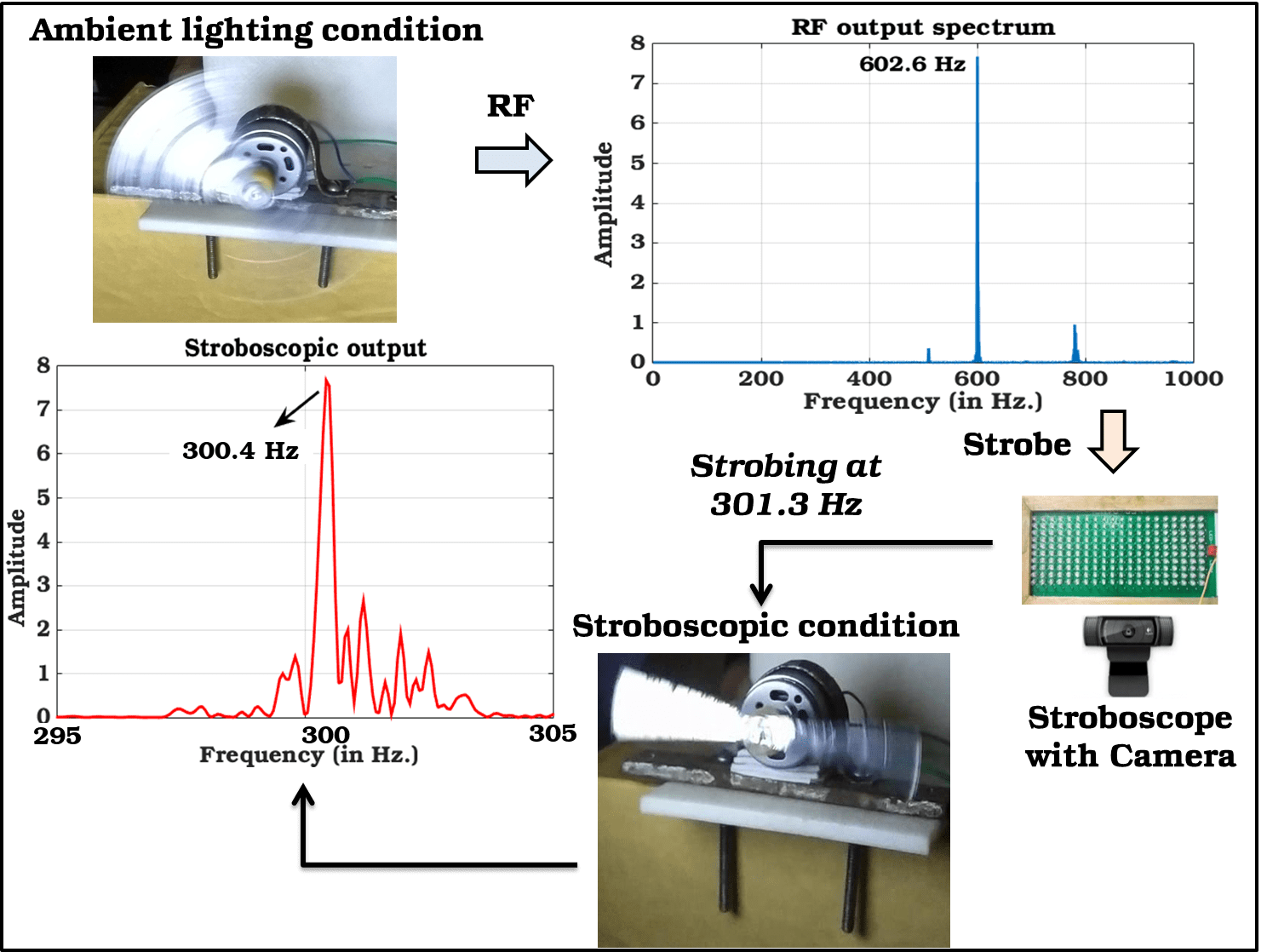}
\caption{Result of rpm measurement of an industrial-graded motor having $18000$ rpm.}
\label{f12.1}
\end{figure}

A similar experiment has been extended for a motor having an rpm of $18000$ with \textit{two} rotating blades (as shown in Fig. \ref{f12.1}). The Doppler-spectrum shows a dominant peak frequency at $602.6$ Hz, thus, the estimated fundamental frequency of rotation is around $301.3$ Hz. Fine-tuning the strobe-frequency through image processing steps, we have noticed a peak-frequency of $0.9$ Hz. Finally, adjusting this value into the stroboscope, the frequency of rotation is found to be $300.4$ Hz that corresponds to $18024$ rpm, subsequently generated $0.13\%$ error detection rate. 

\subsection{Results of Exp 2: Multiple Rotational Motion Detection}
\begin{figure}[!t]
\centering
\includegraphics[width=\columnwidth,height=2.5in]{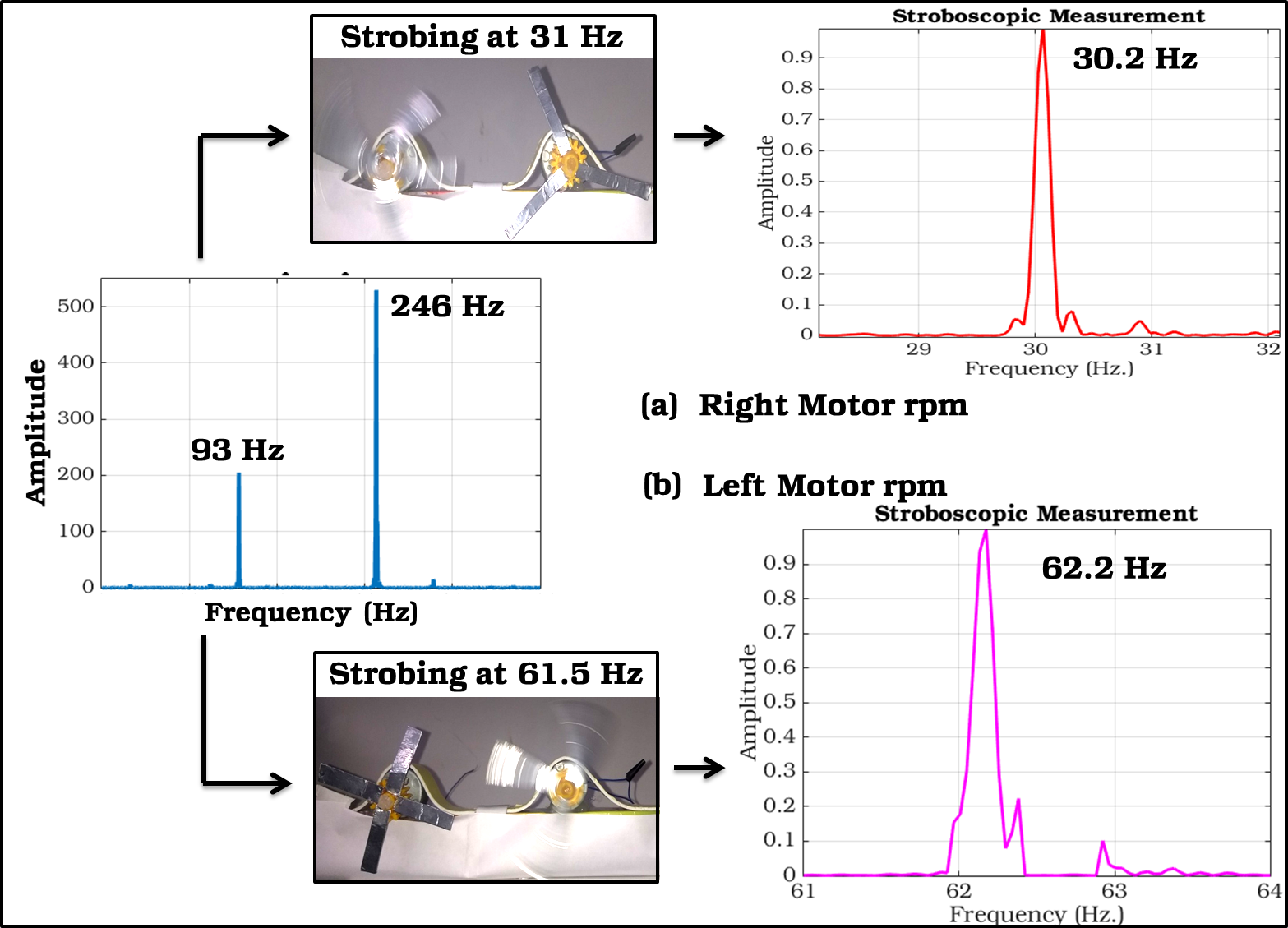}
\caption{Measurement rotational frequencies of multiple motors. (a) Detected frequency of the right motor, (b) Detected frequency of the left motor.}
\label{f13}
\end{figure}

In this experimental study, the proposed system has been applied to sense and localize the rotational frequencies of multiple rotating objects. The Doppler-spectrum provides two district peak frequencies at $93$ and $246$ Hz respectively (as shown in Fig. \ref{f13}). As the two motors contain $3$ and $4$ blades, thus, there might have $4$ feasible combinations of rotational frequencies, i.e. $[23.25, 31, 61.5, 82]$ Hz respectively. 
Scanning through all frequencies sequentially, it is observed that at $31$ Hz strobing frequency, the right motor seems to rotate slowly. Additional improvement of the strobing frequency through the image processing, it is witnessed that the rotational frequency of the right motor is $30.2$ Hz corresponds to $1812$ rpm. 
Furthermore, the rotational frequency of the left motor is obtained as $62.2$ Hz corresponds to $3732$ rpm. Based on the above outcomes, the detection error of the left and right motors will be $0.32\%$ and $0.67\%$ respectively.
\begin{table*}[] 
\centering
\caption{\textcolor{black}{Comparative analysis for the multiple linear motion experimental setup (Fig. \ref{f5}) with the other sensing modalities\newline \textbf{Ground Truth = 113 Hz (left speaker), 141 Hz (right speaker)}}} \vspace{-0.1cm}
\label{table5} \vspace{-0.1cm}
\begin{tabular}{|c|c|c|c|c|c|c|c|}
\hline
\begin{tabular}[c]{@{}c@{}}\textbf{Sensing}\\ \textbf{modality}\end{tabular} & \begin{tabular}[c]{@{}c@{}}\textbf{Detected}\\ \textbf{frequency (Hz)}\end{tabular} & \begin{tabular}[c]{@{}c@{}}\textbf{Detection error}\\ \textbf{$\mu \pm \sigma$ (\%)}\end{tabular} & \begin{tabular}[c]{@{}c@{}}\textbf{Measurement}\\ \textbf{accuracy (\%)}\end{tabular} & \begin{tabular}[c]{@{}c@{}}\textbf{Measurement}\\ \textbf{principle}\end{tabular} & \begin{tabular}[c]{@{}c@{}}\textbf{No. of}\\ \textbf{Sensors}\end{tabular} & \begin{tabular}[c]{@{}c@{}}\textbf{Localization}\\ \textbf{Capability?}\end{tabular} & \begin{tabular}[c]{@{}c@{}}\textbf{Consumed}\\ \textbf{time (sec)}\end{tabular} \\ \hline
Accelerometer \cite{r3.1}                                              & $112.98$, $141.05$ & $0.02 \pm 0.001$ & $99.98$                                               & contact                                                        & $2$        & No                                            & $10$                                                      \\ \hline
LDV \cite{r5}                                                        & $107.35$, $135.36$ & $4.5\pm 1.75$ & $95.5$                                                  & non-contact                                                     & $1$            & Yes                                         & $55$  \\ \hline
RF \cite{r8}             & $112.2$, $140.4$ & $1.13 \pm 0.8$                                    & $98.8$                                                  & non-contact                                                     & 1                      & No                               & $10$                                                 \\ \hline
Camera with strobe \cite{r7.3}               & $112.9$, $141.2$ & $0.23\pm 0.5$                                    & $99.7$                                                  & non-contact                                                     & 1                                                    & Yes & $150$                                               \\ \hline
Proposed                             & $112.95$, $141.12$ & $0.13\pm 0.5$                                                         & $99.9$                                                  & non-contact                                                     & $1$                       & Yes                              & $24$                                                \\ \hline
\end{tabular}
\end{table*} \vspace{-0.5cm}
\subsection{Comparative analysis}
\textcolor{black}{To analyze the efficacies of the proposed technique, we have performed a comparative study with the other well-known vibration measurement systems, for the multiple linear motion experimental setup as shown in Fig. \ref{f5}. The results are shown in Table \ref{table5}. To examine the performance of the sensing modalities, the mean detection error, measurement accuracy, localization capability, and consumed time are selected as the performance indices. Before discussing the results, we would like to mention that each sensing modality has been tested at least $20$ times. Based on those consequences, Table \ref{table5} has been prepared. The mean of the detection error ($e$) is calculated as
\begin{equation}
e = \left[ \frac{1}{N}\sum_{i=1}^{N}\frac{||f_{gt}-f^i_d||}{f_{gt}} \right]\times100\%
\label{DE}
\end{equation}
where $N$ defines the number of times the experiment has performed, $f_{gt}$ is the ground truth frequency of the experiment, $f_d^i$ is the detected frequency at the $i^{th}$ instance. Based on this result, the measurement accuracy (MA) is estimated as
\begin{equation}
MA = \left[ 1 - \frac{e}{100}\right]\times100\%
\label{MA}
\end{equation} 
The consumed time is computed by averaging the execution times of all the experiments for a specific sensor.}

\textcolor{black}{The mean detection error and the measurement accuracy of the accelerometer sensor \cite{r3.1} are around $0.02\%$ and $99.98\%$ respectively. The mean consumed time to detect the vibrational frequencies is around $10$ sec. As this sensor does not have any localization capabilities, in our experiment, we have employed two accelerometers ICs \cite{r3.2} that are attached directly to the diaphragm of the speakers.}

\textcolor{black}{Performing similar experiments with LDV \cite{r5}, it is observed that the detection error and measurement accuracy are around $4.5\%$ and $95.5\%$ respectively. The consumed time is around $55$ sec which is quite large as compared to the other unobtrusive methods.
Executing the same tests with a standalone RF module \cite{r8} or a stroboscopic system \cite{r7.3}, it is witnessed that the detection error and measurement accuracy are less than LDV. However, the consumed time of stroboscope is extremely large, and RF does not have any source localization capability.}

\textcolor{black}{To overcome the above-mentioned shortcomings, we have conducted the corresponding experiment with our proposed solution. It is perceived that the detection error and the measurement accuracy are quite comparable to that of the accelerometer sensor. The consumed time is much less than the standalone stroboscopic system and LDV. Moreover, the system exhibits source localization capabilities autonomously. Therefore, from all of the above consequences, we can claim that the proposed solution could be an alternative in sensing the vibration unobtrusively of machine inspection scenarios.}

\subsection{Results of Exp 3: Wobbling Motion Detection}
\begin{figure}[]
\centering
\includegraphics[width=2.2in,height=1.5in]{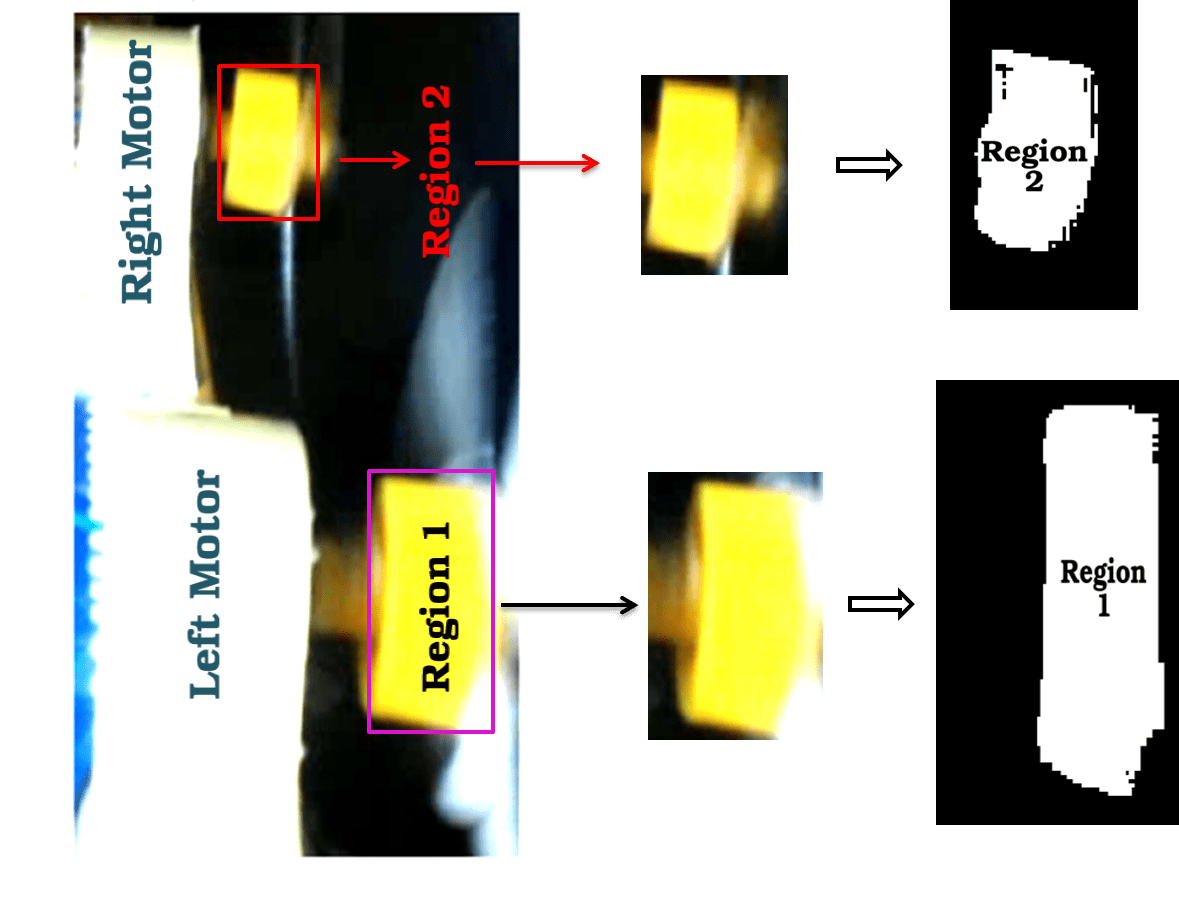}
\caption{Image segmentation in \textit{two}-regions during wobble motion detection.}
\label{f14a}
\end{figure}

\begin{figure}[]
\centering
\includegraphics[width=\columnwidth, height= 2.5in]{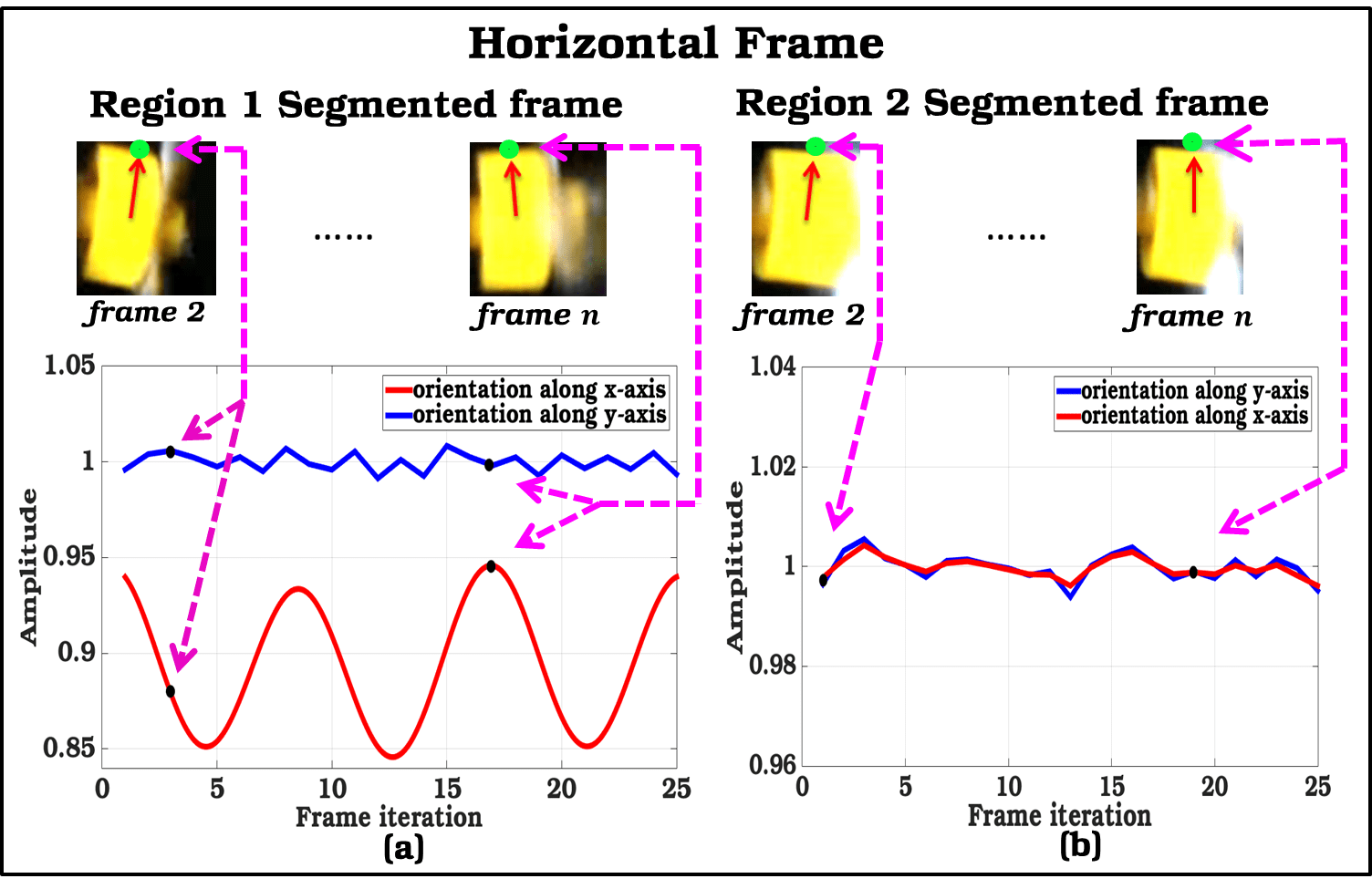}
\caption{\textcolor{black}{The time-variation of the eigenvectors in each of the segmented regions for the horizontal frames.}}
\label{f14b1}
\end{figure}

\begin{figure}[]
\centering
\includegraphics[width=\columnwidth, height=2.5in]{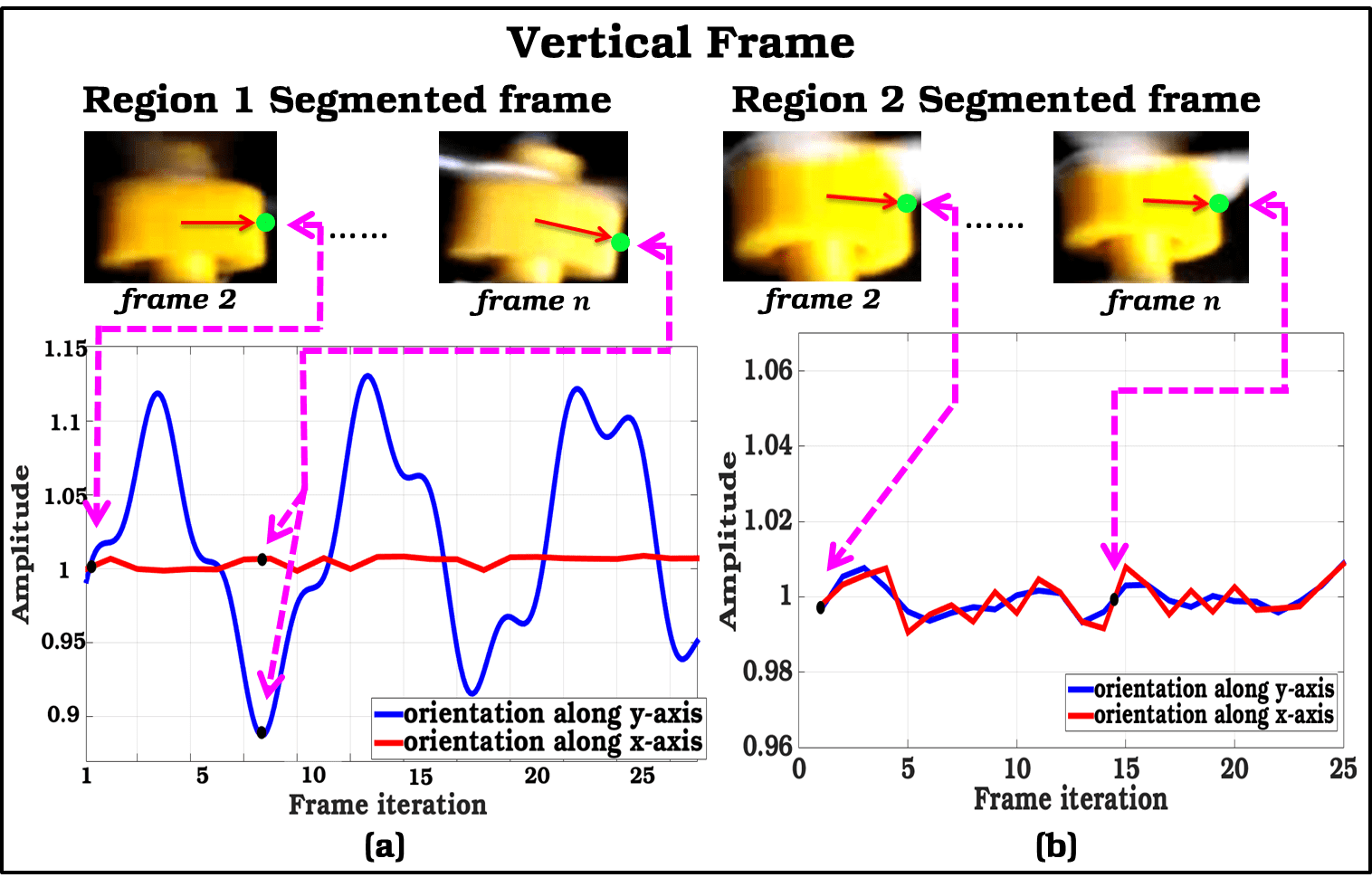}
\caption{\textcolor{black}{The time-variation of the eigenvector in each of the segmented regions in the vertical frames.}}
\label{f14b2}
\end{figure}

\begin{figure}[]
\centering
\includegraphics[width=8.4cm, height=6cm]{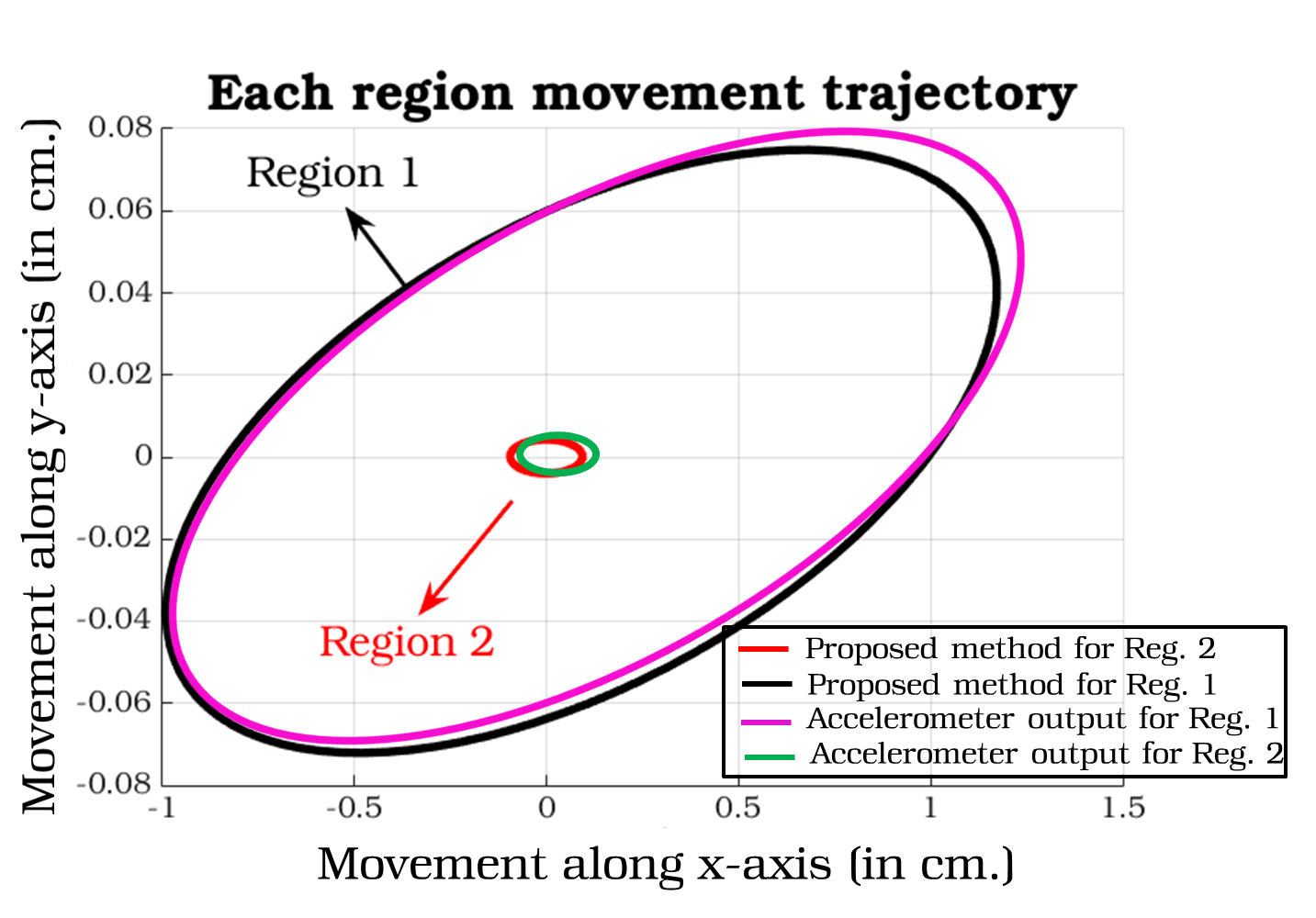}
\caption{\textcolor{black}{Wobble trajectories of the segmented regions in comparison with the horizontally and vertically attached accelerometers.}}
\label{f15}
\end{figure}

The results for the wobbling motion detection (after passing on the radar inputs to the stroboscope for tuning the strobe frequency very close to the motor rpm, as described in Sec. IV) have been depicted in Fig. \ref{f14a}, \ref{f14b1}, \ref{f14b2}, and \ref{f15} respectively. Fig. \ref{f14a} represents a typical segmentation of the object-of-interests in \textit{two} major regions (left and right motors) captured through the camera placed horizontally. The segmented region $1$ and $2$ corresponds to the left and right motors respectively.

On each of the segmented frames (captured through the pair of cameras placed horizontally and vertically, as shown in Fig. \ref{f10c}, we have analyzed the time-variation of the principal components corresponds to the largest eigenvalues of the optical-flow velocity vectors (described in Sec. III-D) while taking the first frame as a reference. The results are shown in Fig. \ref{f14b1} and \ref{f14b2}. Significant variations are witnessed in $x$ (Fig. \ref{f14b1}(a)) and $y$ axis (Fig. \ref{f14b2}(a)) of region $1$, for the frames captured through the horizontal and vertical cameras respectively. However, no such notable variations have been observed for region $2$ (Fig. \ref{f14b1}(b)) and \ref{f14b2}(b)) as it is in a perfectly balanced condition. Furthermore, if we jointly accumulate all the eigenvectors as obtained from the $x$-axis of the horizontal camera frames and $y$-axis of the vertical camera frames for both the regions, we will obtain two different trajectories for region $1$ (marked by \textit{black}-color) and $2$ (marked by \textit{red}-color) respectively as shown in Fig. \ref{f15}. Because of the imbalance in the left-motor, an elliptical trajectory is obtained for the region $1$ (pointed in \textit{black}), whereas, for the perfectly balanced right-motor, a small circular trajectory is obtained for the region $2$ (displayed in \textit{red}). 

To verify the above result, \textit{four} single-axis accelerometer sensors \cite{r3} are attached in the horizontal and vertical directions of the two motors. Fusing the normalized outcomes for both regions, it is seen that the wobble trajectory as obtained from our proposed scheme, coincides with the wobble-trajectory of the accelerometer-based measurement, with a mean-error rate of $1.56\%$ (as shown in Fig. \ref{f15}). 
From this result, we can conclude that our proposed unobtrusive vibration sensing system can estimate the wobbling motion efficaciously.

\textcolor{black}{In all our experiments, we have assumed that the frequency of rotation or vibration of the object-of-interest is fixed and does not vary during the tests. However, in real scenarios, the object-of-interest may have an erroneous periodic movement (i.e. the initial frequency $f_1$ slowly drifts to $f_2$ and again gets back to $f_1$) within the RF measurement time window. In this case, the Doppler-spectrum will capture all the frequency components from $f_1$ to $f_2$. Based on that, the stroboscope will try out all the possible frequencies between $f_1$ to $f_2$ and will look for the freezing effect. As there is a continuous frequency-drift, the stroboscope will not be able to find the intended freezing effect. In this kind of scenario, an alarm could be used to indicate the faulty machine condition.} 

\section{Conclusion}
This paper reports an unobtrusive vibration sensing system, namely \textit{RF-assisted-Strobe} that combines RF radar, stroboscope, and a low-fps camera. The proposed system can be used to measure the vibrational frequencies and the rpm of motors swiftly and precisely. Our extensive experimental outcomes confirm the benefits of this system over the other non-contact vibration measurement system. Moreover, this system can potentially be applied to capture the wobbling motion of a misaligned shaft. Therefore, it could be an attractive solution for affordable sensing in machine condition monitoring or any other scientific applications that requires capturing spatial frequency distribution in a complete unobtrusive fashion. \textcolor{black}{In future, we would like to study the performance of the system in an anechoic chamber with faulty machines that generate complex periodic rotation.}


%

\ifCLASSOPTIONcaptionsoff
  \newpage
\fi



%
%
%

\bibliographystyle{IEEEtran}
\bibliography{IEEEabrv,dib}

\end{document}